\documentclass[useAMS,referee,usenatbib]{biom}
\def\bSig\mathbf{\Sigma}

\usepackage{amsmath}
\usepackage{amsfonts}
\usepackage{amsxtra}
\usepackage{MnSymbol}
\usepackage{bbm}
\usepackage{hyperref}
\usepackage[dvips]{graphicx}
\usepackage{wrapfig}
\usepackage{boxedminipage}
\usepackage{color}

\title{Joint estimation of genetic and parent-of-origin effects using RNA-seq data from human}

\author
{Vasyl Zhabotynsky \\
 Department of Biostatistics, University of North Carolina at Chapel Hill,
 Chapel Hill, North Carolina, U.S.A
 \and
Wei Sun\emailx{wsun@fredhutch.org}\\
Public Health Sciences Division, Fred Hutchinson Cancer Research Center, Seattle, Washington, U.S.A.
\and
Kaoru Inoue \\
Department of Pharmacology, University of North Carolina at Chapel Hill, 
Chapel Hill, North Carolina, U.S.A.
\and
Terry Magnuson \\
Department of Genetics, University of North Carolina at Chapel Hill, 
Chapel Hill, North Carolina, U.S.A. 
\and
Mauro Calabrese \\
Department of Pharmacology, University of North Carolina at Chapel Hill, 
Chapel Hill, North Carolina, U.S.A.
}

\begin{document}





\pagerange{\pageref{firstpage}--\pageref{lastpage}} 
\pubyear{2016}
\artmonth{October}




\label{firstpage}


\begin{abstract}

RNA sequencing allows one to study allelic imbalance of gene expression, which may be due to genetic factors or genomic imprinting. It is desirable to model both genetic and parent-of-origin effects simultaneously to avoid confounding and to improve the power to detect either effect. In a study of experimental cross, separation of genetic and parent-of-origin effects can be achieved by studying reciprocal cross of two inbred strains. In contrast, this task is much more challenging for an outbred population such as human population. To address this challenge, we propose a new framework to combine experimental strategies and novel statistical methods. Specifically, we propose to collect genotype data from family trios as well as RNA-seq data from the children of family trios. We have developed a new statistical method to estimate both genetic and parent-of-origin effects from such data sets. We demonstrated this approach by studying 30 trios of HapMap samples. Our results support some of previous finding of imprinted genes and also recover new candidate imprinted genes. 
\end{abstract}

%



\maketitle


%

\section{Introduction}
\label{s:intro}
By combining RNA-seq data and phased genotype of a diploid genome, one can estimate allele-specific expression (ASE). Allelic imbalance of gene expression (i.e., the ASE of the two alleles are different) may be controlled by genetic factors or genomic imprinting. Genetic loci that affect gene expression variation are often referred to as gene expression quantitative trait loci (eQTLs). Sophisticated statistical methods have been developed  to identify eQTLs using both ASE and total expression \citep{sun2012asSeq,sun2013eqtl,hu2015proper}. Genomic imprinting is a phenomenon in which the expression of a gene depends whether an allele has maternal or paternal origin. Some genes are (completely) imprinted in the sense that they are exclusively expressed from one of the two parental alleles. However, recent studies have shown that many genes are partially imprinted, i.e., both alleles are expressed while the allele from one parent has higher expression the other allele \citep{crowley2015analyses}. Although only a small percentage of human genes are (completely or partially) imprinted, studying of imprinting is of great  importance for at least two reasons. First, the imprinted genes play important roles in many common diseases such as obesity or psychiatric disorders \citep{peters2014role}. Second, genomic imprinting is also a great model to study epigenetic regulation of gene expression \citep{barlow2011genomic}. \\

It is important to jointly estimate genetic and parent-of-origin effects on allelic imbalance of gene expression in order to avoid confounding of these two factors and to improve the power to detect either effect. This goal can be achieved by studying RNA-seq data from reciprocal cross of two inbred strains, e.g., F1 mouse population \citep{zou2014novel,crowley2015analyses}. Specifically, denote two inbred strains by $A$ and $B$,  a reciprocal cross refers to two crosses $A \times B$ and $B \times A$. Following standard naming convention, a cross is written as maternal strain $\times$ paternal strain, and thus in a cross $A \times B$, $A$ is the maternal strain and $B$ is the paternal strain. To make it more clear, we write the two crosses as $A^{m} \times B^{p}$ and $B^{m} \times A^{p}$, where the superscript indicates maternal or paternal origin. In one cross $A^{m} \times B^{p}$, if the expression of $A^{m}$ allele is larger than $B^{p}$ allele, it is not clear whether it is due to genetic effect (i.e., $A$ allele to higher expression) or parent-of-origin effect (i.e., the maternal allele has higher expression). However, given a reciprocal cross, such confounding can be resolved.  \\

In an outbred population such as the human population, separation of genetic and parent-of-origin effect becomes much more challenging. Previous studies have employed genotyping arrays to assess allelic imbalance using both RNA and genomic DNA (gDNA) samples and identify potential imprinted genes if the degrees of allelic imbalance is significantly different between RNA and gDNA samples. Those potential imprinted genes are then screened using family pedigree  \citep{morcos2011genome} or experimental validation \citep{barbaux2012genome}. A more direct approach is to collect RNA-seq data from the children of family trios (a family trio includes both parents and one child) and collect genotype data from all three family members of each trio. The genotype  can be phased and imputed to a larger number of SNPs. Then ASE can be quantified based on phased genotype data. In this work, we developed a new statistical method to separate genetic and parent-of-origin effect while jointly modeling ASE and total read count (TReC) across a group of family trios.

\section{Method}
\label{s:method}
We assume genetic effects can be captured by a small number of \textit{cis}-acting eQTLs. Here we define \textit{cis}-eQTLs as those eQTLs that influence allelic-imbalance of gene expression \citep{sun2013eqtl}. This assumption can be justified by the following rationales. First, by definition, only \textit{cis}-acting eQTLs are relevant for ASE study. Second, the number of \textit{cis}-acting eQTLs is limited because most \textit{cis}-acting eQTLs are local eQTLs that are close to the gene of interest, and due to linkage dis-equilibrium (LD) structure around a gene, the number of independent \textit{cis}-eQTLs of a gene is relatively small. In the following, we assume there is only one \textit{cis}-eQTL to simplify the discussion. Our method can be easily extended to the cases with multiple \textit{cis}-eQTLs. 

\subsection{Allele specific read counts (ASReCs)}

To simplify the notation, we assume the haplotypes connecting a candidate eQTL and the gene of interest are known. Let $h_{k1}$ and $h_{k2}$ be the two haplotypes of a gene of interest in the $k$-th individual (for $k=1, ... , K$), and let the  $N_{k1}$ and $N_{k2}$ be the allele-specific read counts of these haplotypes $h_{k1}$ and $h_{k2}$, respectively.  Let $N_k = N_{k1} + N_{k2}$. Denote the two alleles of the candidate eQTL as $A_1$ and $A_2$, and denote its genotype in the $k$-th individual as $g_k$. We assign different meanings to genotypes $A_1A_2$ and $A_2A_1$ such that $A_1A_2$ means haplotypes $h_{k1}$ and $h_{k2}$ harbor the $A_1$ and $A_2$ alleles, respectively, and $A_2A_1$ means haplotypes $h_{k1}$ and $h_{k2}$ harbor the $A_2$ and $A_1$, respectively. We model $N_{k1}$ by a beta-binomial distribution (denoted by $f_{BB}$):
\begin{eqnarray}
N_{k1} \sim f_{BB}(N_{k1}; N_{k}, p_k, \phi), \ \ \  \log\left[{p_k}/({1 - p_k})\right] = b_0 z_k + b_1 x_k,
\label{eq:01}
\end{eqnarray}
where $\phi$ is over-dispersion parameter, and $z_k$ and $x_k$ are defined as follows:
\[
z_k = \left\{ 
\begin{array}{l l}
  0  &  \quad \text{if $g_k =  A_iA_i, \ i=1 \textrm{ or } 2$ }\\
  1  &  \quad \text{if $g_k =  A_2A_1$ }\\
  -1  &  \quad \text{if $g_k =  A_1A_2$; }\\
\end{array} \right. 
\]
\[
x_k = \left\{ 
\begin{array}{l l}
  1  &  \quad \text{if haplotype $h_{k1}$ is inherited from the paternal allele, }\\
  -1  &  \quad \text{if haplotype $h_{k1}$ is inherited from the maternal allele. }\\
\end{array} \right. 
\]

\subsection{Total read counts (TReCs)}
The TReC of the gene of interest in the $k$-th individual, denoted by $T_k$, is modeled by a negative binomial distribution with mean $\mu_k$ and over-dispersion parameter $\varphi$, denoted by $f_{NB}(T_k; \mu_k, \varphi)$. We can write the mean structure for negative binomial distribution as: 
\begin{eqnarray}
\log(\mu_{k}) = \gamma_0 + \beta_{\kappa} \log(\kappa_k) + \sum_{u=1}^p \beta_{u} c_{ku} +  \eta_k,
\label{eqn:mukAA}
\end{eqnarray}
where $\beta_{u}$, $u=1, ..., p$, is the regression coefficient for the $u$-th covariate (e.g., age, gender, batch effects). $\eta_k$ is defined as:
\[ \eta_k =
  \begin{cases}
    0    &  \quad \text{if } g_k = A_1A_1 \\
    \log\left\{1+\exp(b_0 + x_k b_1)\right\} - \log\left\{1 + \exp(x_k b_1)\right\}   &  \quad \text{if } g_k = A_1A_2 \textrm{ or } A_2A_1 \\
    b_0   &  \quad \text{if } g_k = A_2A_2 \\
  \end{cases}
\]
Note that the additive genetic and parent-of-origin effects are parametrized by $b_0$ and $b_1$, which are the same as the $b_0$ and $b_1$ in equation~(\ref{eq:01}) for allele-specific read counts. 
 
\subsection{Joint likelihood}
The joint likelihood of total read count (TReC) and allele-specific read count (ASReC) is 
\begin{eqnarray}\label{e:jt}
  \L({\Theta}) = 
   \prod_{k=1}^{K}f_{BB}(N_{k1}, N_{k}; b_0, b_1, \phi)f_{NB}(T_k; b_0,b_1,\varphi, \gamma,\beta_{\kappa},\beta_1,...\beta_p), 
 \end{eqnarray}
where $\Theta=(b_0,b_1,\phi,\varphi,\gamma,\beta_{\kappa},\beta_1,...\beta_p)$. In this model we assume common genetic and parent-of-origin effect for TReC and ASReC. We test the genetic and parent-of-origin effects by testing whether $b_0$ or $b_1$ equals to 0, respectively, with the likelihood ratio test.  

\subsection{Optimization Algorithm}

For a given initial values for non-linear terms ($\phi$, $\varphi$, $b_0$, $b_1$), our optimization algorithm includes the following steps. \\

Step 1: Optimize linear terms given the initial values of non-linear terms:
\begin{eqnarray*}
\beta_{r+1} = \beta_{r}+(X'W_{r}X)^{-1}(X'W_{r}k_{r}), \ 
\textrm{diag}(W_{r})=\frac{\mu_{r}}{1+\phi_r^{-1}\mu_{r}} \quad \textrm{and} \quad  k_r=\frac{y_r-\mu_r}{\mu_r}, 
\end{eqnarray*}
where $W_r$ is a diagonal matrix.\\

Step 2: Iteratively estimate $b_0$ and $b_1$. Note that the following two steps are redundant, but improves the robustness of the algorithm. 
\begin{enumerate}
  \item Optimize $b_0$ and $b_1$ together using BFGS method. 
  \item Optimize $b_0$ and $b_1$ separately using Brent algorithm. \\
\end{enumerate}

Step 3: optimize over-dispersion parameter $\log(\phi)$ and $\log(\varphi)$ separately using Brent algorithm. For stability we limit range of over-dispersion to be between $10^{-4}$ and $10^4$.\\

Step 4: If likelihood change is larger than a small number $\epsilon$, go to step 1, otherwise finish the estimation process. 

\section{Simulations}
We simulated ASE and TReC from the model described by joint likelihood in equation~(\ref{e:jt}). We chose the smallest sample size to be 32, which is similar to the sample size we have for real data analysis. To demonstrate the asymptotic properties, we also simulated data of larger sample sizes of 64, 128, and 256. We set the over-dispersion parameters for beta-binomial ($\varphi$) and negative binomial ($\phi$) distributions to be 1/4, and 4/3, respectively, which are fairly typical in real data. We further scale the expected counts so that mean total read count was about 250. We set the proportion of counts that are allele-specific to be 10\% which is reasonably close to observed value in our real data.
\begin{figure}[h]
\begin{center}
   \includegraphics[width=1.00\textwidth]{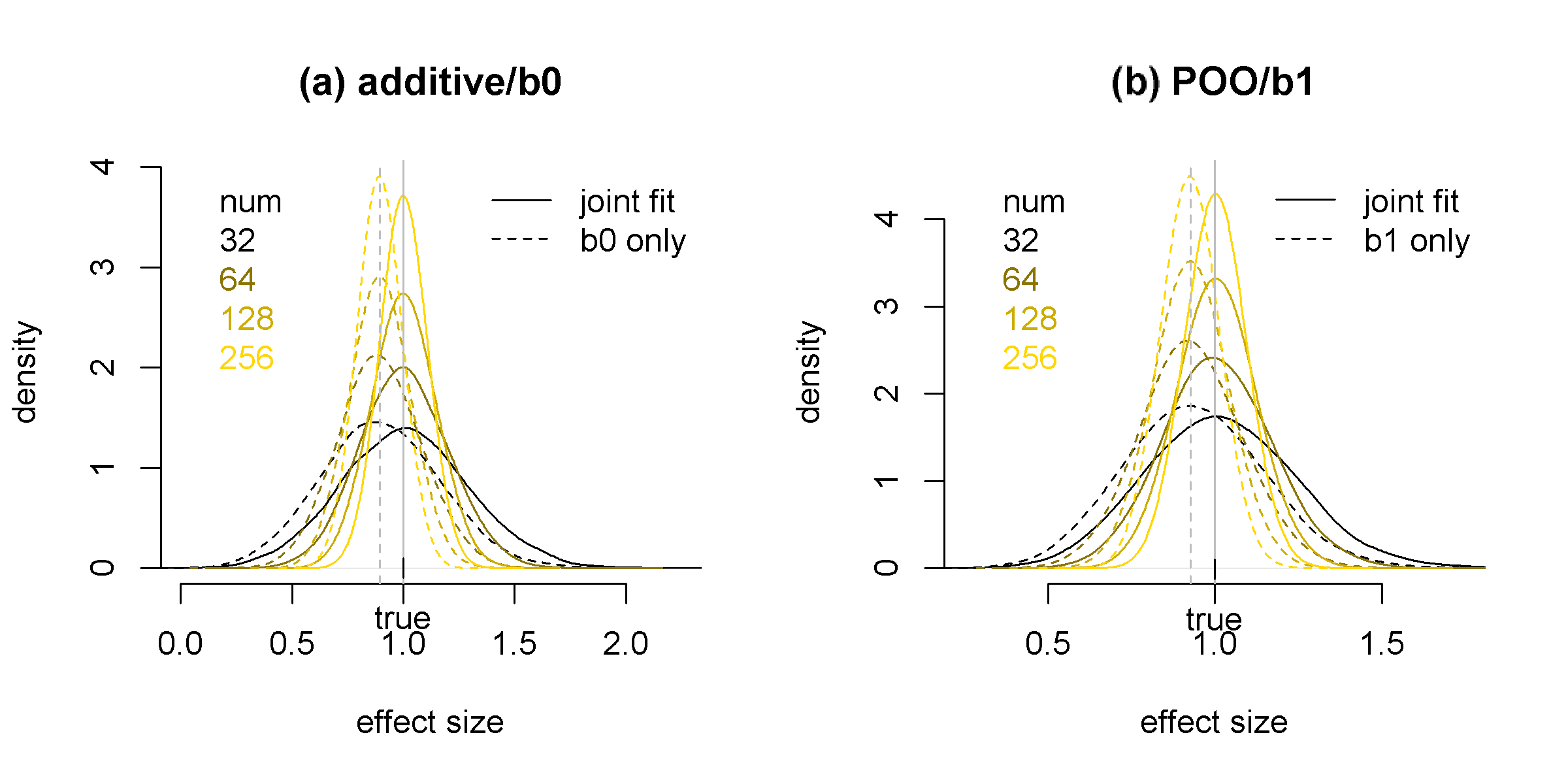}
   \caption{Fitting the joint model (a) additive genetic effect and (b) parent-of-origin effect for various sample sizes. Left legend describes number of simulated individuals, solid line densities show model estimating both additive and parent-of-origin effect jointly, while dashed densities show what happens when only one effect is fitted (i.e. additive only for left panel and parent-of-origin only for right panel)}
   \label{f:jtunbi}
\end{center}
\end{figure}

Our simulation results show that the estimates of $b_0$ and $b_1$ from our method are unbiased. In contrast, if a model is fit with only genetic or parent-of-origin effect while in fact both effects are present, the parameter estimates have significant bias 
(Figure \ref{f:jtunbi}).  

We also compared the type I error and the power of our joint model versus naive fit using $R/glm.nb$ function to fit Negative-Binomial model using total read counts only and $R/vglm$ function to fit Beta-Binomial model using allele specific counts only. As shown in Table 1, the simple models don't control type I error as well as the joint model and that the joint model has higher power to detect both genetic and parent-of-origin effects.

\begin{table}
 \caption{Power Analysis }
 \label{t:one}
  \centering
  \begin{tabular}{ll | lll | lll}
  \hline
  \multicolumn{2}{|c|}{Parameters} & \multicolumn{3}{|c|}{Genetic} &  \multicolumn{3}{|c|}{Parent of Origin}\\ \cline{3-5} 
 $$  & $$   &  Joint    &   Negative  &  Beta   &  Joint  &  Negative  &  Beta \\
 $b_0$  & $b_1$   &  Model   &  Binomial  &  Binomial   &  Model  &  Binomial  &  Binomial \\
  \hline
0.00 & 0.00 & 0.06 & 0.11 & 0.07 & 0.06 & 0.12 & 0.07\\
0.13 & 0.13 & 0.09 & 0.12 & 0.08 & 0.11 & 0.11 & 0.10\\
0.25 & 0.25 & 0.19 & 0.14 & 0.11 & 0.27 & 0.11 & 0.18\\
0.50 & 0.50 & 0.53 & 0.21 & 0.21 & 0.72 & 0.12 & 0.46\\
0.75 & 0.75 & 0.83 & 0.32 & 0.33 & 0.96 & 0.14 & 0.76\\
1.50 & 1.50 & 1.00 & 0.76 & 0.49 & 1.00 & 0.39 & 0.93\\
  \hline
  \end{tabular}
\end{table}

\begin{figure}[h]
\begin{center}
   \includegraphics[width=1.00\textwidth]{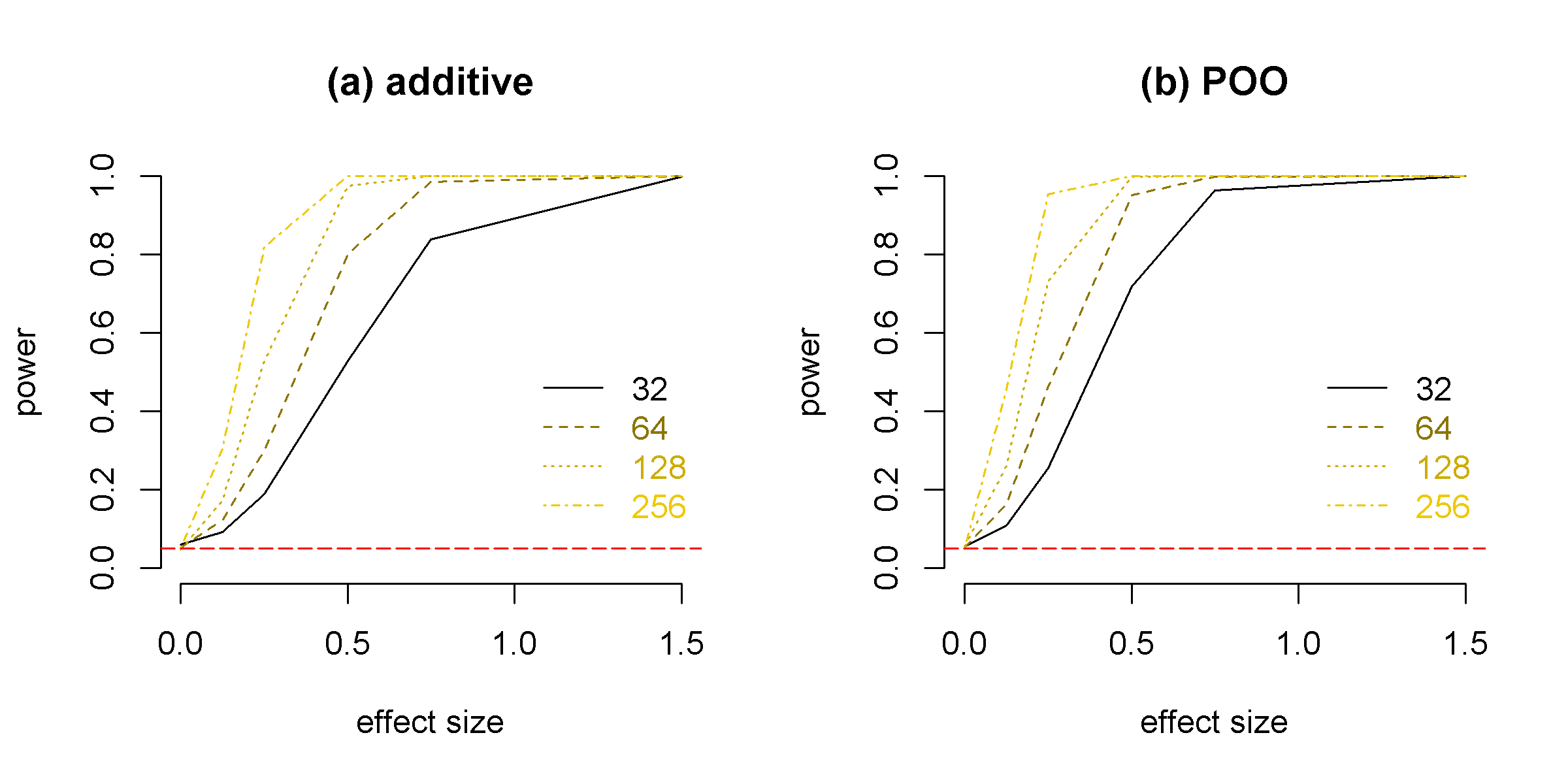}
   \caption{The statistical power of our joint model to detect (a) additive genetic effect or (b) parent-of-origin effects for various sample sizes. The horizontal red dash line indicates the p-value cutoff 0.05 to declare statistical significance. }
   \label{f:power}
\end{center}
\end{figure}

The statistical power of the joint model is illustrates in Figure~\ref{f:power}. Even at sample size of 32, our method has around 80\% of power to detect either genetic or parent-of-origin effect at two-fold change, which corresponds to effect size $\log(2) = 0.693$. The power to detect parent-of-origin effect is higher than the power to detect genetic effect. This is because the ASE of all samples can be used to quantify parent-of-origin effect (i.e., comparing paternal vs. maternal allele). In contrast, ASE can be used to quantify genetic effect only if the genotype of the candidate eQTL is heterozygous, which only happen in a subset of the samples. 

The computational time of our method scales well with the sample size. The time per gene remains under a second for sample sizes under 200 individuals: see Figure 6 in Supplementary Materials .

\section{Application}
\subsection{Data collection}
We collected RNA-seq data from 30 HapMap CEU (Caucasian) samples (15 males + 15 females). All of these samples are children of family trios, where the genotypes/haplotypes of three family members have been reported in the HapMap project, and their lymphoblastoid cell lines are available through Coriell (\url{http://www.coriell.org/}). For most of these samples, the RNA reads were 150 bp paired-end reads, with an additional smaller run of 75 bp paired-end reads. The median of the total number of reads for these 30 samples is about 20 million. These reads were mapped to hg38 human reference genome using Tophat2. Hapmap project genotyped about 3.9 million SNPs for these 30 trios. We phased and imputed the genotypes of these 30 trios using shapeit2 \citep{shapeit2} and impute2 \citep{howie2012fast}, against 1000 Genome reference panel containing 2,504 individuals with $\sim$82 million SNPs. Finally, based on phased and imputed genotype, we extracted allele-specific reads (i.e., those RNA-seq reads that overlap with at least one heterozygous SNP), and counted the number of allele-specific reads for each haplotype of a gene. 

\subsection{Identification of candidate \textit{cis}-eQTLs}
The parents of these 30 HapMap family trios are part of the samples included in the 1000 Genomes project \citep{10002012integrated}. To improve the power and precision for eQTL mapping, we first mapped \textit{cis}-acting eQTLs using the Caucasian samples of the 1000 Genome Project. Then given the eQTLs, we jointly estimated genetic and parent-of-origin effects in the 30 children of HapMap trios. We downloaded fastq files for 462 samples from the Geuvadis consortium \citep{lappalainen2013transcriptome}. We mapped all the reads to hg38 reference, and then performed the same process of phasing and imputation of the genotypes of these samples. Then we calculated total read counts and allele-specific read counts per gene and per sample. After filtering out non-European samples as well as samples with low quality in terms of systematic bias of allele-specific expression (see section A.9 in Supplementary Materials for details), we performed eQTL mapping for 227 samples using the TReCASE method \citep{hu2015proper}. For each gene, we identified one candidate \textit{cis}-acting eQTL by selecting the one with the smallest p-value. Note that some genes do not have any \textit{cis}-acting eQTL because all the nearby SNPs show certain evidence of \textit{trans}-acting eQTL, i.e., the eQTL genotype is associated with total expression, but not allele-specific expression. The eQTL effect sizes estimated from these 227 samples are consistent with the estimates from the 30 children of HapMap trios (see section B.1 in Supplementary Materials for details).

\subsection{Identification of imprinted genes}
We used our method to estimate genetic (eQTL) and imprinting effects for 12,386 genes in the 30 children of the family trios. These genes were selected because they had enough expression in the 30 samples and they had at least one candidate \textit{cis}-eQTL based on eQTL mapping results from the 227 samples of 1000 Genome Project. For the negative binomial model of total read count (TReC), we fit our model with additional covariates to capture batch effect (the RNA-seq data were collected through 3 batches with 10 samples per batch) and read-depth. No additional covariate is needed for the analysis of allele-specific read counts (ASReC) because our model for ASE compared the expression of one allele versus another allele, and thus the effects of such covariates are canceled. We found 16 genes with significant imprinting effects at q-value cutoff 0.05 (Table~\ref{tb:poo-genes}). 
\begin{table}
 \caption{A list of genes with potential parent-of-origin effects. We classify genes with q-value between 0.05 and 0.25 into three groups. $^1$: missed cutoff of q-value 0.05 due to low count; $^2$ missed cutoff due to smaller effect size. }
\label{t:pootable}
  \centering 
  \begin{tabular}{lllll}
\hline
ID  &  Name  &  q-value  &  Expression  &  Is Known \\
\hline
ENSG00000269821 & KCNQ1OT1 & 1.2e-08 & paternally & yes\\
ENSG00000204186 & ZDBF2 & 9.6e-08 & paternally & yes\\
ENSG00000167981 & ZNF597 & 2.9e-07 & maternally & yes\\
ENSG00000185513 & L3MBTL1 & 5.1e-06 & paternally & yes\\
ENSG00000177432 & NAP1L5 & 6.6e-06 & paternally & yes\\
ENSG00000242265 & PEG10 & 6.9e-06 & paternally & yes\\
ENSG00000257151 & PWAR6 & 5.8e-05 & paternally & no\\
ENSG00000224078 & SNHG14 & 8.2e-05 & paternally & no\\
ENSG00000130844 & ZNF331 & 1.2e-04 & paternally & no\\
ENSG00000261069 & RP11-701H24.4 & 3.8e-04 & paternally & no\\
ENSG00000225806 & RP1-309F20.3 & 3.8e-04 & paternally & no\\
ENSG00000122390 & NAA60 & 4.3e-04 & maternally & yes\\
ENSG00000128739 & SNRPN & 3.6e-03 & paternally & yes\\
ENSG00000100138 & SNU13 & 4.4e-03 & paternally & no\\
ENSG00000145945 & FAM50B & 2.5e-02 & paternally & yes\\
ENSG00000101898 & MCTS2P & 3.7e-02 & paternally & yes\\
\hline
ENSG00000279192$^{1}$  & PWAR5         & 8.6e-02 & paternally & no\\
ENSG00000174851$^{2}$  & YIF1A         & 9.9e-02 & paternally & no\\
ENSG00000182109$^{1}$  & RP11-69E11.4  & 1.3e-01 & paternally & no\\
ENSG00000171847$^{1}$ & FAM90A1       & 1.5e-01 & maternally & no\\
ENSG00000082781$^{1}$  & ITGB5         & 2e-01   & paternally & no\\
ENSG00000178057$^{1}$  & NDUFAF3       & 2e-01   & maternally & no\\
ENSG00000254319$^{1}$  & RP11-134O21.1 & 2e-01   & paternally & no\\
ENSG00000253633$^{2}$  & KB-1980E6.3   & 2e-01   & paternally & no\\
ENSG00000111678$^{1}$  & C12orf57      & 2e-01   & maternally & no\\
ENSG00000101160      & CTSZ          & 2e-01   & maternally & no\\
ENSG00000054967$^{2}$ & RELT          & 2.1e-01 & paternally & no\\
ENSG00000126226$^{2}$  & PCID2         & 2.1e-01 & maternally & no\\
ENSG00000135709      & KIAA0513      & 2.1e-01 & maternally & no\\
ENSG00000175643      & RMI2          & 2.1e-01 & maternally & no\\
ENSG00000262155$^{1}$ & RP11-266L9.5  & 2.2e-01 & paternally & no\\
\hline
  \end{tabular}
  \label{tb:poo-genes}
\end{table}
Out of these 16 genes, 10 were known to be imprinted from previous studies and 6 were novel candidates for imprinting. For 14 of these 16 genes, the paternal allele had higher expression than the maternal allele. For all of those 10 known imprinted genes, our estimates of allelic imbalance agree with what were reported before. At a more liberal cutoff of q-value 0.25, we identified 15 additional candidates of imprinted genes. By manually examining the expression pattern of these 15 genes, we suggest that 9 of them missed the q-value cutoff 0.05 due to power - either because parent-of-origin effect was smaller (3 genes) or the number of allele-specific reads was small (6 genes). The other 6 are likely false discoveries, because there is no apparent pattern of allelic imbalance for one parental allele. 

\begin{figure}[h]
\begin{center}
   \includegraphics[width=0.9\textwidth]{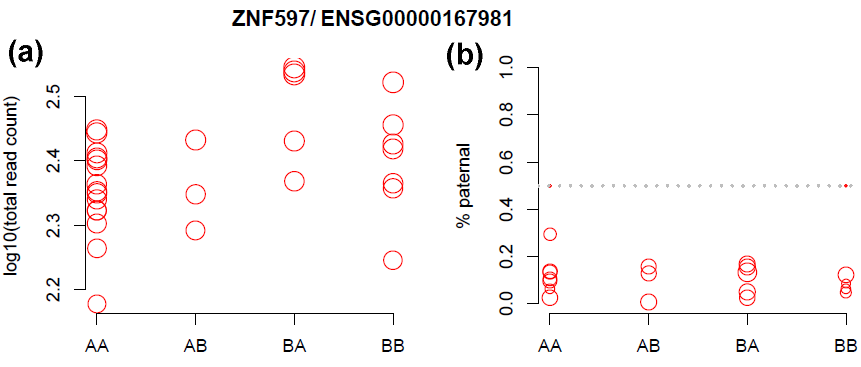}
   \caption{ZNF497 - a maternally expressed gene. (a) normalized total read counts ($\log_{10}$ scale), (b) percent of paternally expressed reads. Reads are classified into four categories by their genotype, assuming that first recorded genotype is maternal. Size of circle reflects the scale of log read counts.}
   \label{f:imprex}
\end{center}
\end{figure}

We illustrate the read count data of a clearly imprinted gene, ZNF497, which has higher expression on maternal allele (Figure~\ref{f:imprex}). 
The imprinting effect can be observed from both TReC and ASReC. We denote the genotype such as the first allele is maternal allele, i.e., genotype AB means A and B are from maternal and paternal allele, respectively. For TReC, the two groups with genotype AA and AB have similar expression because they share the same maternal allele and maternal allele has higher expression. Similarly, the two groups BA and BB have similar expression.  The effect of imprinting is more apparent in allele-specific reads where we observe the proportion of reads from paternal allele is far below 50\%.

We selected known imprinted genes based on the list reported by \cite{morison2005census} or those genes recorded in the Geneimprint database \citep{geneimprint2016}. There is a total of 90 genes classified as imprinted. Out of those, 32 were expressed in our dataset and had a candidate eQTL. Of those 32 genes 10 were found to be significant (q-value $<$ 0.05) by our method. For several genes (such as RB1, KCNQ1, PEG3, and PLAGL1), we observed signal of imprinting, but the signal was too weak to produce significant q-value. In general, though, we observed that even for insignificant results, those with relatively smaller q-value tend to have estimated imprinting direction matched with the reported imprinting direction.  

\subsection{Locations of genes with parent-of-origin effect}
\begin{figure}[h]
\begin{center}
   \includegraphics[width=1\textwidth]{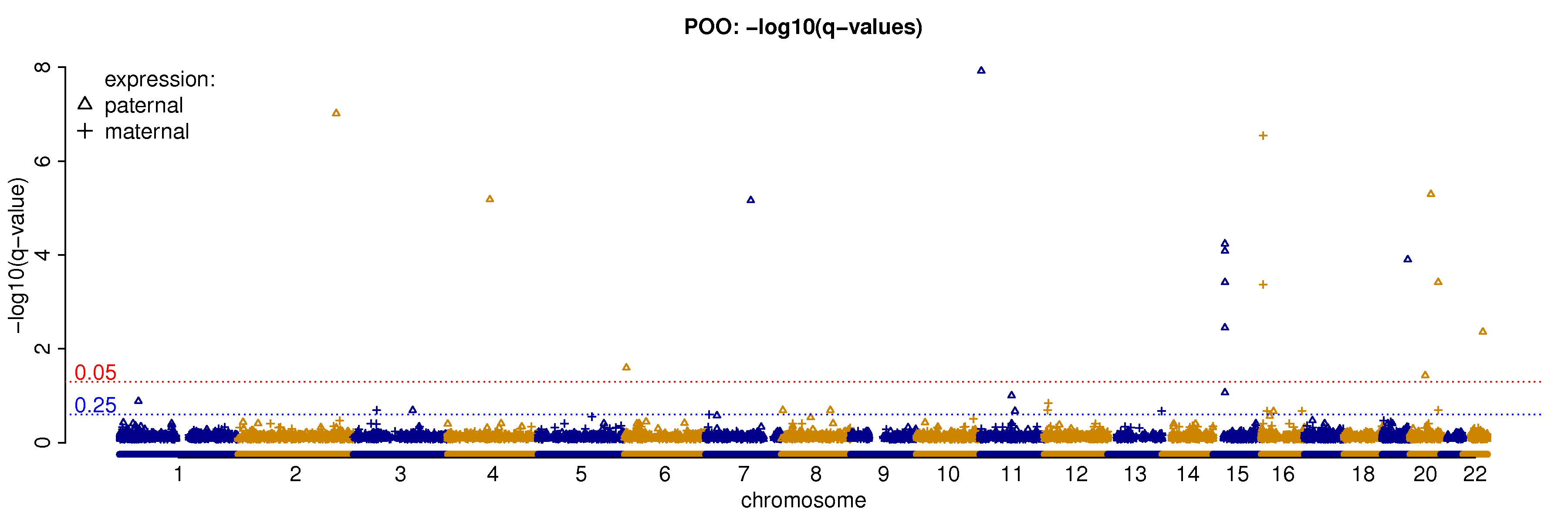}
   \caption{Positions of discovered parent of origin effects.}
   \label{f:poopos} 
\end{center}
\end{figure}
The imprinted genes in mouse genome tend to form clusters. It is interesting to check whether this is true for human genome. Based on the findings from our study, the imprinted genes on chromosomes 15 and 16 form two clusters (Figure~\ref{f:poopos} and Table~\ref{t:poobychr}). However, the total number of imprinted genes is small and thus it is difficult to formally assess the statistical significance of such spatial clusters. On the other hand, while most imprinted genes have higher expression on the paternal allele, there is some variation at chromosome level. We can test whether the proportion of paternally imprinted (or equivalently maternally imprinted) genes is uniform across all chromosomes. In order to have enough genes for testing, we define potential imprinted genes at very liberal q-value cutoffs 0.25 or 0.5, and only consider the chromosomes with at least 2 imprinted genes. At q-value cutoff $0.25$ and $0.5$, the p-values for Fisher's exact test are $0.02$ and $0.04$, respectively, suggesting that genes within one chromosome have more consistent pattern of parent-of-origin expression than expected by chance. 
Next, for each chromosome, we assess whether the proportion of paternally imprinted genes of this chromosome is different from all the other chromosomes. We performed a test only if there are at least 5 imprinted genes in a chromosome (Table~\ref{t:poobychr}). These formal tests confirm the clusters at chromosome 16 and 15, at imprinting q-value cutoff 0.25 and 0.5, respectively; although at the q-value cutoff 0.5, the p-value for chromosome 15 is not significant after multiple testing correction. 

\begin{table}
 \caption{POO genes found by chromosome}\label{t:poobychr}
  \centering 
  \begin{tabular}{lllll}
  \multicolumn{1}{|c|}{} & \multicolumn{2}{|c|}{q-value $< 0.25$} &  \multicolumn{2}{|c|}{q-value $< 0.5$}\\ 
\hline
chr & pat/tot  &  p-value  &   pat/tot  &  p-value  \\
\hline
1 & 1/1 &           &7/11&0.36\\
2 & 1/1 &           &5/11&1.00\\
3 & 1/2 &           &2/6 &0.68\\
4 & 1/1 &           &6/7 &0.06\\
5 & 0/0 &           &1/8 &0.06\\
6 & 1/1 &           &5/8 &0.49\\
7 & 1/1 &           &3/5 &0.68\\
8 & 2/2 &           &5/8 &0.49\\
9 & 0/0 &           &1/1 &\\
10 & 0/0 &         &2/6 &0.68\\
11 & 3/3 &         &5/10&1.00\\
12 & 0/2 &         &1/8 &0.06\\
13 & 0/1 &         &0/4 &\\
14 & 0/0 &         &5/10&1.00\\
15 & 5/5 &0.29  &7/8 &0.03\\
16 & 1/5 &0.02  &3/11&0.21\\
17 & 0/0 &         &5/8 &0.49\\
18 & 0/0 &         &2/3 &\\
19 & 1/1 &         &5/11&1.00\\
20 & 3/4 &         &5/8 &0.49\\
21 & 0/0 &         &0/0 &\\
22 & 1/1 &         &1/3 &\\
\hline
  \end{tabular}
\end{table}

\section{Discussion}
We developed a systematic approach to jointly estimate genetic and parent-of-origin effect in human. Our results recovered about one third of known imprinted genes, and if we excluded genes with low expression and included weaker, but non-contradicting cases, there was a good overall consistency. 
None of the genes classified as ``predicted imprinting'' instead of ``imprinting'' in Geneimprint database were detected as high confidence in our short list. One possible reason is that most of these genes  were selected based on a screening paper \citep{luedi2007computational} and were false positives in that work. We also noted that for these ``predicted imprinting'' genes, the proportion of genes with predominantly paternal expression roughly 50\%, while this proportion is about 68\% (61/90) for known imprinting genes. Several imprinted genes that we found are non-coding RNAs, such as RP11-701H24.4 and RP1-309F20.3, which warrants further studies to elucidate the functional consequence if imprinted non-coding RNAs. 

The raw RNA-seq data were submitted to NCI SRA database. The R package of our method will be submitted to CRAN and posted at \url{http://research.fhcrc.org/sun/en/software.html}. 




\backmatter

\section*{Supplementary Materials}

Web Appendix, referenced in Section 2-4, is attached below.\vspace*{-8pt}

\section{RNA-seq data of 30 children of HapMap family trios}
\vspace{2ex}

\subsection{Data collection and mapping}
RNA-seq data were collected from 30 children of HapMap family trios (15 males + 15 females) in five runs: 10 individuals sequenced in November 2014 with 150 bp paired-end reads (batch 1), 10 sequenced twice in December 2014 with 150 bp paired-end reads (batch 2), and 10 sequenced twice in January 2015 with most of the reads sequenced with 150bp paired-end reads and another smaller run with 75 bp paired-end reads (batch 3). RNA-seq reads were mapped to hg38 reference assembly using TopHat v2 \cite{trapnell2009tophat}, filtered with the following criteria: $\leq$ 3 read mismatches, read gap length $\leq$ 3, read edit distance $\leq$ 3, and read realign edit distance equals to 0.\\

\subsection{Quality control}
\begin{figure}[h]
\begin{center}
   \includegraphics[width=0.9\textwidth]{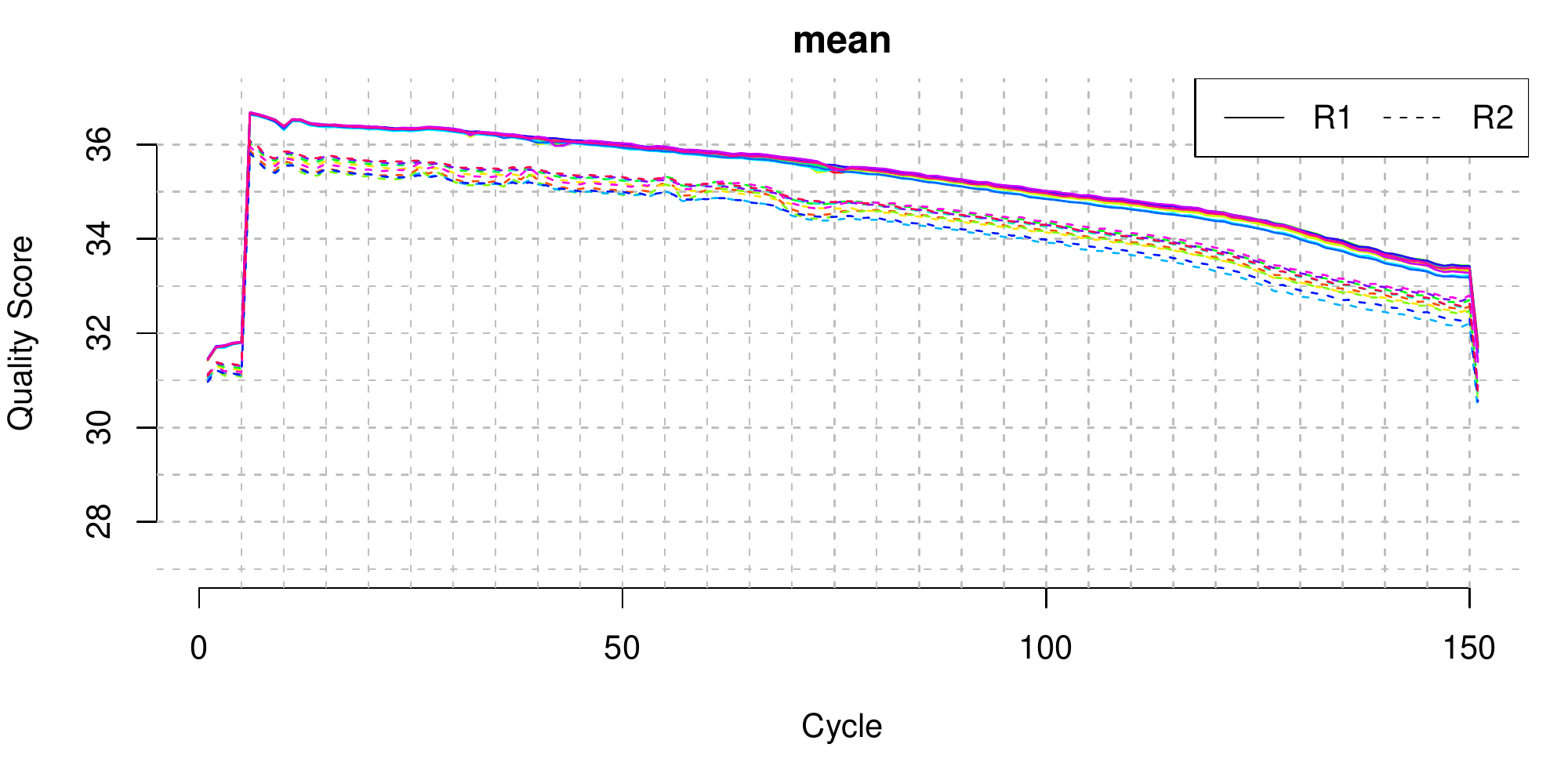}
   \caption{Mean read quality by the base for 10 samples of the first batch (November 2014). R1 and R2 denote first and second end of the reads for each sample}
   \label{f:qual} 
\end{center}
\end{figure}

We further filtered RNA-seq reads with average base quality $\geq$ 30, mapping quality $\geq$ 20, and keeping only those uniquely mapped reads, by the prepareBAM function from R package \texttt{asSeq}. In a typical sample, most of the RNA-seq reads pass these filters. Figure~\ref{f:qual} illustrates mean base quality for the first batch (November 2014). \\

\subsection{Matching genotype coordinates to hg19}
During phasing and imputation we worked with the reference panel mapped to hg19. Therefore the available genotype data was mapped to hg19 coordinates using liftover tool \cite{rhead2009ucsc}. After phasing and imputation was done, the resulting SNPs were mapped to hg38 coordinates, because the RNAseq data were mapped to hg38. 

\subsection{Phasing}
For this phasing step and the next imputation step, we used the 1000 Genome reference panel \cite{howie2011genotype} (as of summer 2015) containing 2,504  individuals with $\sim$82 million SNPs. For the 90 individuals from 30 family trios, phasing was done using shapeit v.2. Family information of the 30 trios was used to improve the phasing accuracy.  We started with $\sim$3.9 million SNPs of these 30 trios obtained from HapMap project, $\sim$1.3 million of which were filtered out due to following reasons:  $\sim$30\% of them were absent in the reference panel, $\sim$70\% had mismatching information. Effective size of the population was set to the suggested value \texttt{--effective-size 20000} and random seed was set to $1234567$.\\

\subsection{Imputation}
The results of phasing can be directly used as input for impute2, with imputation done using the latest reference available at impute2 web page \cite{howie2009flexible}. Imputation was done by splitting the genome into blocks of 5 Mb (no more than 7 Mb according to the instruction). We also used the same population size option as the one used in phasing step (\texttt{-Ne 20,000}), other options used: \texttt{-align\_by\_maf\_g} and \texttt{-seed 12345}.\\

\subsection{Creating the lists of heterozygous SNPs for each individual}
We obtained the list of heterozygous SNPs for each sample from the results of imputation. The main output file (other files have additional extensions to the name as \texttt{\_haps}, \texttt{\_allele\_probs} etc) has 3 columns with genotype probabilities per SNP. This file represents probability of observing $G=0, G=1, G=2$ respectively. We selected heterozygous locations (i.e. with high probability of $G=1$) to output phased genotypes of these locations. \\

\subsection{Finding allele specific reads}
Once we got a list of heterozygous SNPs, we extracted allele-specific reads for two haplotypes using R function \texttt{extractASReads} from R package \texttt{asSeq} \cite{sun2012asSeq}. The extracted reads were saved in two additional bam files. We assigned the parent of origin of the two haplotypes based on how well they match the paternal and maternal haplotypes. \\

\subsection{Gene level summarization}
\begin{figure}[h]
\begin{center}
   \includegraphics[width=.55\textwidth]{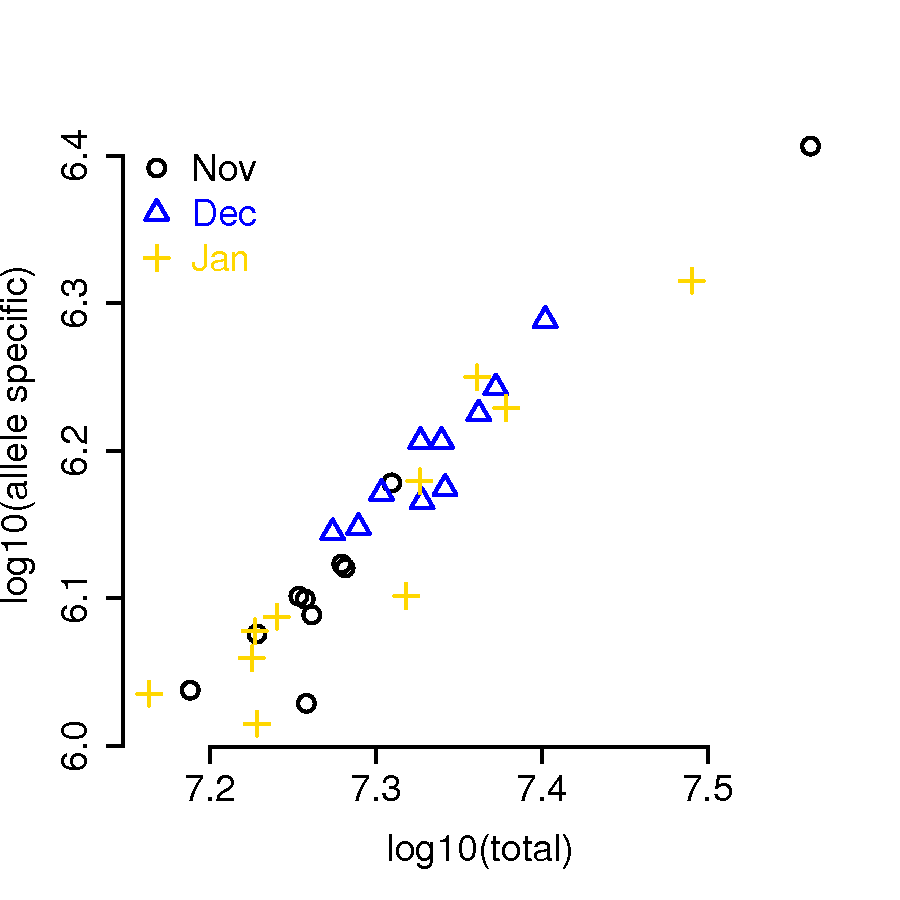}
   \caption{Relationship between total read counts and allele specific read counts by sample.}
   \label{f:aseperc} 
\end{center}
\end{figure}

Finally, we summarized main bam file as well as the parent-specific bam files using R function \texttt{summarizeOverlaps} from R package \texttt{GenomicAlignments} \cite{lawrence2013software} with option ``mode = Union'' and ``yieldSize = 1,000,000''. Across all the 30 samples, the percentage of reads being allele-specific are consistent (Figure~\ref{f:aseperc}). Approximately $6.3\%$ of the reads can be classified as allele specific. Comparing with percent of allele-specific counts from unrelated population we conclude that having a family structure allowed to increase the percentage of allele-specific reads by about $0.5\%$. Additionally we can see that the proportion of reads with conflicting haplotype information is quite low in all samples (Table ~\ref{t:cnt}).\\

\begin{table}
\small
 \caption{Gene level count summary. Haplotype N denotes reads with conflicting SNP information}
 \vspace{1ex}
   \label{t:cnt} 
  \centering 
  \begin{tabular}{llllllllll}
  \hline
$id$&$Haplotype 1$&$Haplotype 2$&$Haplotype N$&$Total$&$\%ASE$&$\%Hap.N$&$Month$\\
  \hline

12386&531735&508232&7551&15424873   &6.7\%&0.05\%&Nov\\
07349&1170915&1216844&17937&36471875&6.5\%&0.05\%&Nov\\
10855&549410&553054&9853&16920863   &6.5\%&0.06\%&Nov\\
10856&614655&619313&10258&19030106  &6.5\%&0.05\%&Nov\\
10857&573787&589056&9100&17933576   &6.5\%&0.05\%&Nov\\
10860&526271&518571&8804&18122047   &5.8\%&0.05\%&Nov\\
10864&601558&612843&9985&19123476   &6.4\%&0.05\%&Nov\\
12335&726865&664370&10215&20396307  &6.8\%&0.05\%&Nov\\
12344&575986&567731&8277&18256805   &6.3\%&0.05\%&Nov\\
12376&580297&581345&9352&18093944   &6.4\%&0.05\%&Nov\\
10831&687568&683270&7854&21967310   &6.2\%&0.04\%&Dec\\
10837&808867&810355&9829&23568960   &6.9\%&0.04\%&Dec\\
10839&646210&638965&6174&19480573   &6.6\%&0.03\%&Dec\\
10840&770903&767830&9354&23016699   &6.7\%&0.04\%&Dec\\
10845&700172&683873&7286&20112291   &6.9\%&0.04\%&Dec\\
10861&645650&655743&8762&18791340   &6.9\%&0.05\%&Dec\\
12707&691861&770957&7185&21231491   &6.9\%&0.03\%&Dec\\
12740&886096&898011&10693&25241951  &7.1\%&0.04\%&Dec\\
12766&727328&751223&8373&21859216   &6.8\%&0.04\%&Dec\\
12767&686563&678695&7300&21279339   &6.4\%&0.03\%&Dec\\
10847&110414&108922&1086&5236664    &4.2\%&0.02\%&Jan\\
10851&104600&103799&1099&4431665    &4.7\%&0.02\%&Jan\\
12752&117984&125960&1333&6632011    &3.7\%&0.02\%&Jan\\
12753&196200&200245&1841&9581273    &4.1\%&0.02\%&Jan\\
12801&101874&99410&908&5246099      &3.8\%&0.02\%&Jan\\
12802&172863&169508&1659&7640589    &4.5\%&0.02\%&Jan\\
12818&146227&145745&1612&6461376    &4.5\%&0.02\%&Jan\\
12832&154141&165334&1668&6917632    &4.6\%&0.02\%&Jan\\
12864&117679&118731&1038&5389998    &4.4\%&0.02\%&Jan\\
12877&111281&109520&1012&5203301    &4.2\%&0.02\%&Jan\\
  \hline
  \end{tabular}
\end{table}

\subsection{Processing of 462 lymphoblastoid cell lines from the 1000 Genomes samples}
\begin{figure}[h]
\begin{center}
   \includegraphics[width=0.8\textwidth]{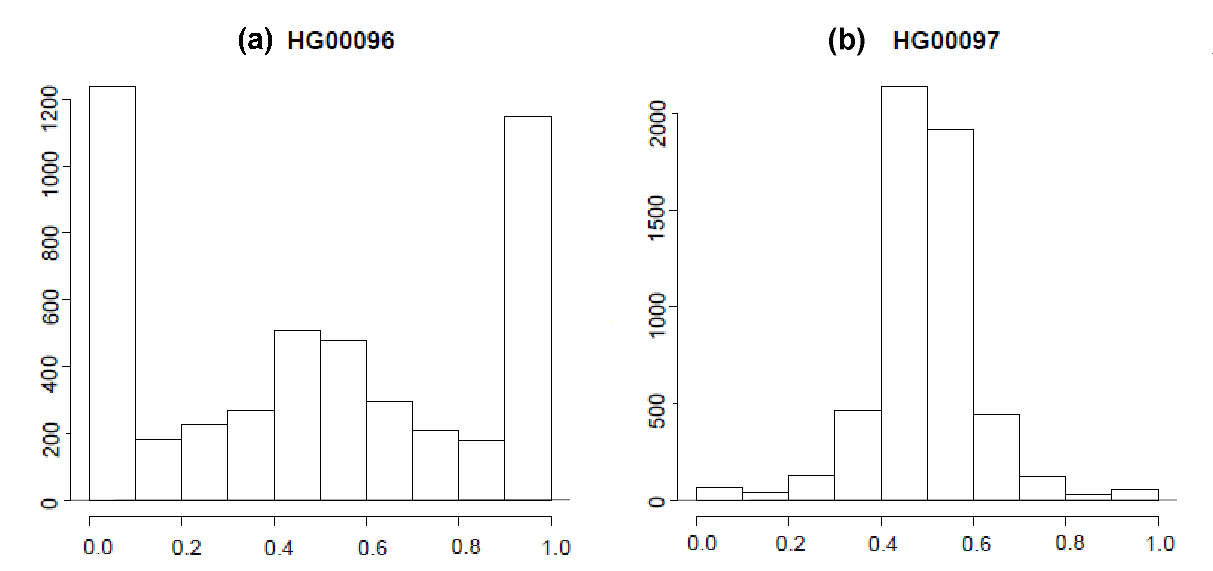}
   \caption{The distribution of the proportion of reads from haplotype 1 across all genes for (a) a bad sample and (b) a good sample.}
   \label{f:confl} 
\end{center}
\end{figure}

A similar process as in previous subsections was applied on RNA-seq data of 462 lymphoblastoid cell lines from the 1000 Genomes \cite{lappalainen2013transcriptome}: unrelated human lymphoblastoid cell line samples from the CEU, FIN, GBR, TSI and YRI populations (see E-GEUV-1 data link below) were sequenced with Illumina HiSeq2000 platform, with paired-end 75-bp reads. We mapped these reads to the hg38 reference with the same options as for original 30 individuals. To produce allele specific reads we started with 2,123 individuals of African or European descent with $\sim$2.2 million SNPs per individual produced by the Omni Array data originally from \cite{delaneau2012linear}. Since African sub-population is quite different from Western European sub-population, we removed this sub-population and also removed samples, for which we observed many genes having extreme proportions of reads attributed to one haplotype - this is likely due to the fact of having a notable proportion of SNPs with problematic genotypes. Figure~\ref{f:confl} shows typical samples we classify bad/good samples. A principal component analysis of total expression for the remaining 227 European samples suggested that western European individuals (CEU), which are the most comparable to our 30 samples, are also comparable to other sub-European populations. Thus we used these 227 samples for eQTL mapping.\\

\section{eQTL analysis for 227 Caucasian samples from 1000 genome project}
\vspace{2ex}

\subsection{Comparing 227 Caucasian samples versus our 30 HapMap samples}
\begin{figure}[h]
\begin{center}
   \includegraphics[width=0.8\textwidth]{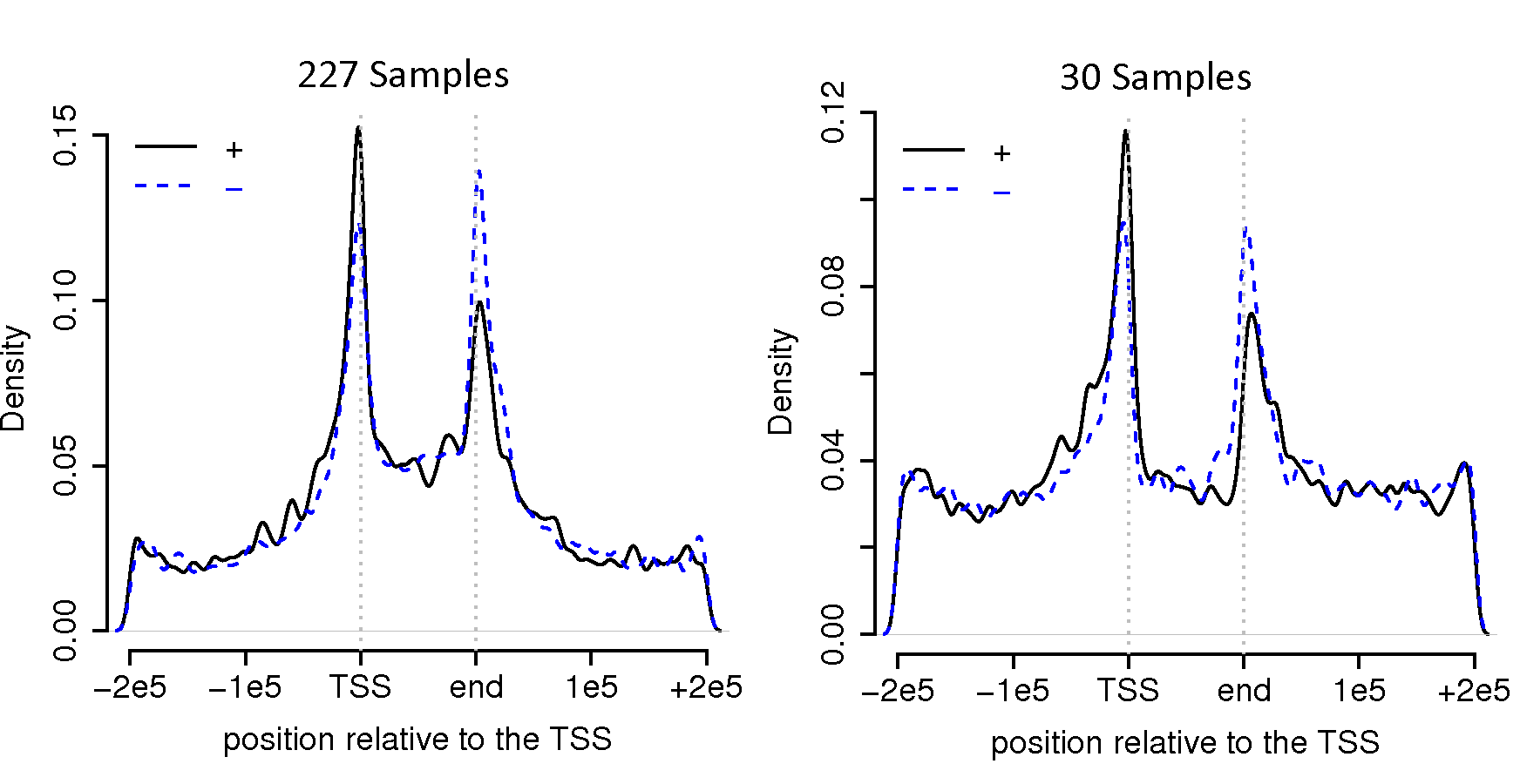}
   \caption{Distance to transcription start site: left - 227 samples, right - 30 children. Genes on positive or negative strands were plot separately. }
   \label{f:tssdist} 
\end{center}
\end{figure}

\begin{figure}[h]
\begin{center}
   \includegraphics[width=0.45\textwidth]{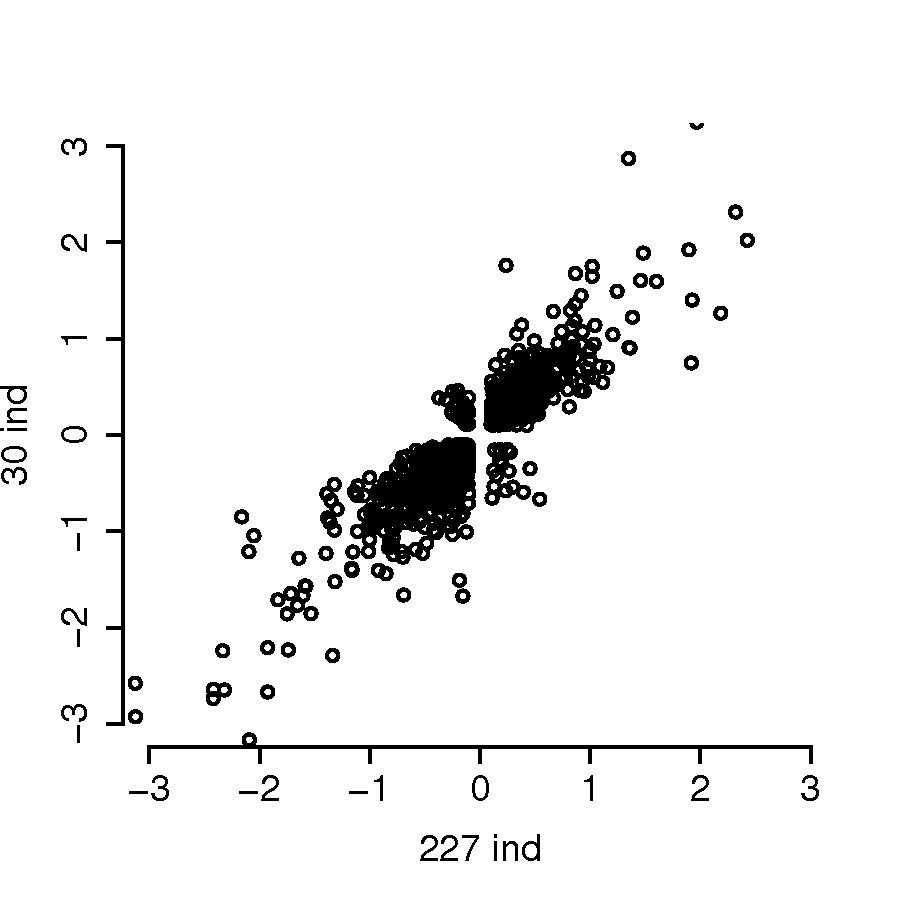}
   \caption{Comparison of additive genetic effect estimates using 227 samples from 1000 genome project versus 30 children from family trios.}
   \label{f:betas} 
\end{center}
\end{figure}

We mapped \textit{cis}-eQTLs using TReCASE model for the 227 Caucasian samples as well as our 30 samples (30 children of family trios). Figure~\ref{f:tssdist} shows the location distribution of these \textit{cis}-eQTLs relative to the transcription starting site of the corresponding genes. The distributions derived from the 227 Caucasian samples and our 30 samples are very similar. 
In addition, the estimates of genetic effects  using 227 individuals vs. 30 children at the same SNP (considering only SNPs declared significant at $\alpha=0.05$ were also consistent (Figure \ref{f:betas}).\\

\clearpage

\section{Additional simulations}
\vspace{2ex}
\subsection{Timing}
\begin{figure}[h]
\begin{center}
   \includegraphics[width=.85\textwidth]{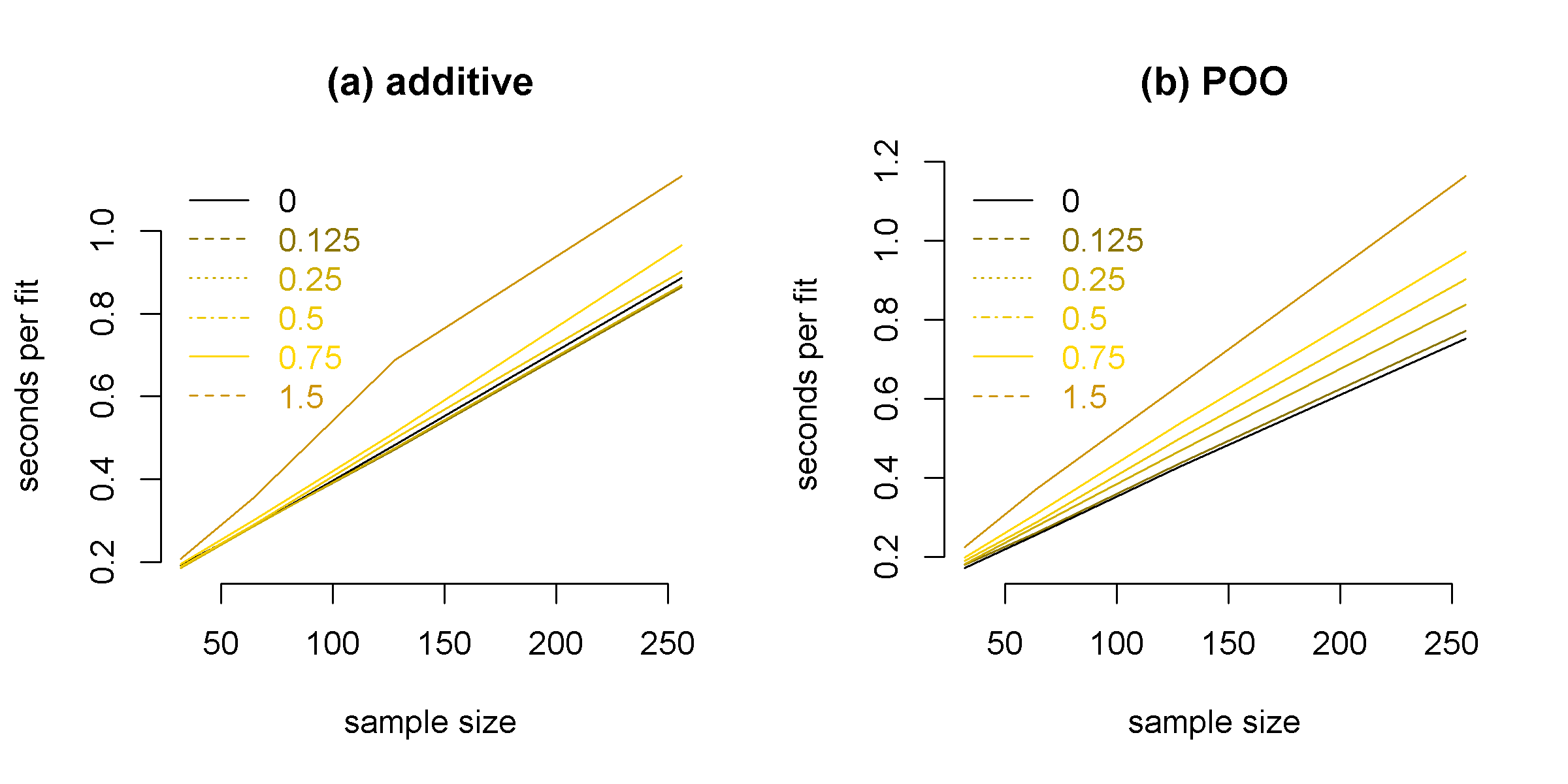}
   \caption{Time (in seconds) to fit full and two short models per gene for various sample sizes and several effect sizes.  (a) Varying additive genetic effect with parent-of-origin effect fixed at 0.5. (b) Varying parent-of-origin effect with additive genetic fixed at 0.5.} 
   \label{f:timing} 
\end{center}
\end{figure}

We define one model fit as the following. First fit a TReC only model with $b_0$ to obtain reasonable initial estimate of $b_0$, and then fitting full TReCASE model estimating $b_0$ and $b_1$, as well as two short models for additive genetic effect or parent-of-origin effect only. The computational time needed for one model fit (i.e., for one gene) is linearly dependent on sample size and increases for parameters being farther away from zero (Figure~\ref{f:timing}). For the sample size of around 200 the computation time is about 1 second per gene.

\section{Software and Data}
\vspace{2ex}

RNA-seq data from 1000G samples (E-GEUV-1 data) \cite{lappalainen2013transcriptome} is available from 
\url{http://www.ebi.ac.uk/arrayexpress/experiments/E-GEUV-1/samples/}\\ 

\noindent Liftover \cite{rhead2009ucsc}: \url{https://genome.ucsc.edu/util.html}\\

\noindent Shapeit v2 \cite{shapeit2}: \url{https://mathgen.stats.ox.ac.uk/genetics_software/shapeit/shapeit.html}\\

\noindent IMPUTE v2 \cite{howie2009flexible}: \url{http://www.stats.ox.ac.uk/~marchini/software/gwas/gwas.html}\\

\noindent 1000 Genome Reference Panel \cite{howie2011genotype}: \url{https://mathgen.stats.ox.ac.uk/impute/impute_v2.html#reference}\\

\noindent TopHat V2 \cite{trapnell2009tophat}: \url{https://ccb.jhu.edu/software/tophat/index.shtml}\\

\noindent \texttt{R/GenomicAlignments} \cite{lawrence2013software}: \url{https://bioconductor.org/packages/release/bioc/html/GenomicAlignments.html}\\

\noindent \texttt{R/asSeq} \cite{sun2012asSeq}: \url{http://research.fhcrc.org/sun/en/software/asSeq.html} \\

\clearpage

\backmatter

\section*{Acknowledgements}
\vspace*{-8pt}

This research is supported in part by NIH grant GM105785.

\clearpage
\bibliographystyle{biom}
\bibliography{imprinting_ref}

\begin{thebibliography}{}

\bibitem[\protect\citeauthoryear{{1000 Genomes Project Consortium}}{{1000
  Genomes Project Consortium}}{2012}]{10002012integrated}
{1000 Genomes Project Consortium} (2012).
\newblock An integrated map of genetic variation from 1,092 human genomes.
\newblock {\em Nature} {\bf 491,} 56--65.

\bibitem[\protect\citeauthoryear{Barbaux, Gascoin-Lachambre, Buffat, Monnier,
  Mondon, Tonanny, Pinard, Auer, Bessi{\`e}res, Barlier, et~al\mbox{.}}{Barbaux
  et~al.}{2012}]{barbaux2012genome}
Barbaux, S., Gascoin-Lachambre, G., Buffat, C., Monnier, P., Mondon, F.,
  Tonanny, M.-B., Pinard, A., Auer, J., Bessi{\`e}res, B., Barlier, A., et~al.
  (2012).
\newblock A genome-wide approach reveals novel imprinted genes expressed in the
  human placenta.
\newblock {\em Epigenetics} {\bf 7,} 1079--1090.

\bibitem[\protect\citeauthoryear{Barlow}{Barlow}{2011}]{barlow2011genomic}
Barlow, D.~P. (2011).
\newblock Genomic imprinting: a mammalian epigenetic discovery model.
\newblock {\em Annual review of genetics} {\bf 45,} 379--403.

\bibitem[\protect\citeauthoryear{Crowley, Zhabotynsky, Sun, Huang, Pakatci,
  Kim, Wang, Morgan, Calaway, Aylor, et~al\mbox{.}}{Crowley
  et~al.}{2015}]{crowley2015analyses}
Crowley, J.~J., Zhabotynsky, V., Sun, W., Huang, S., Pakatci, I.~K., Kim, Y.,
  Wang, J.~R., Morgan, A.~P., Calaway, J.~D., Aylor, D.~L., et~al. (2015).
\newblock Analyses of allele-specific gene expression in highly divergent mouse
  crosses identifies pervasive allelic imbalance.
\newblock {\em Nature genetics} {\bf 47,} 353--360.

\bibitem[\protect\citeauthoryear{Delaneau, Marchini, Consortium,
  et~al\mbox{.}}{Delaneau et~al.}{2014}]{shapeit2}
Delaneau, O., Marchini, J., Consortium, . G.~P., et~al. (2014).
\newblock Integrating sequence and array data to create an improved 1000
  genomes project haplotype reference panel.
\newblock {\em Nature communications} {\bf 5,}.

\bibitem[\protect\citeauthoryear{Delaneau, Marchini, and Zagury}{Delaneau
  et~al.}{2012}]{delaneau2012linear}
Delaneau, O., Marchini, J., and Zagury, J.-F. (2012).
\newblock A linear complexity phasing method for thousands of genomes.
\newblock {\em Nature methods} {\bf 9,} 179--181.

\bibitem[\protect\citeauthoryear{Howie, Fuchsberger, Stephens, Marchini, and
  Abecasis}{Howie et~al.}{2012}]{howie2012fast}
Howie, B., Fuchsberger, C., Stephens, M., Marchini, J., and Abecasis, G.~R.
  (2012).
\newblock Fast and accurate genotype imputation in genome-wide association
  studies through pre-phasing.
\newblock {\em Nature genetics} {\bf 44,} 955--959.

\bibitem[\protect\citeauthoryear{Howie, Marchini, and Stephens}{Howie
  et~al.}{2011}]{howie2011genotype}
Howie, B., Marchini, J., and Stephens, M. (2011).
\newblock Genotype imputation with thousands of genomes.
\newblock {\em G3: Genes, Genomes, Genetics} {\bf 1,} 457--470.

\bibitem[\protect\citeauthoryear{Howie, Donnelly, and Marchini}{Howie
  et~al.}{2009}]{howie2009flexible}
Howie, B.~N., Donnelly, P., and Marchini, J. (2009).
\newblock A flexible and accurate genotype imputation method for the next
  generation of genome-wide association studies.
\newblock {\em PLoS Genet} {\bf 5,} e1000529.

\bibitem[\protect\citeauthoryear{Hu, Sun, Tzeng, and Perou}{Hu
  et~al.}{2015}]{hu2015proper}
Hu, Y.-J., Sun, W., Tzeng, J.-Y., and Perou, C.~M. (2015).
\newblock Proper use of allele-specific expression improves statistical power
  for cis-{eQTL} mapping with {RNA}-seq data.
\newblock {\em Journal of the American Statistical Association} {\bf 110,}
  962--974.

\bibitem[\protect\citeauthoryear{Jirtle}{Jirtle}{2016}]{geneimprint2016}
Jirtle, R.~L. (2016).
\newblock Gene imprint, imprinted gene database (internet)Available at
  \url{http://www.geneimprint.com/site/genes-by-species}.

\bibitem[\protect\citeauthoryear{Lappalainen, Sammeth, Friedl{\"a}nder,
  AC‘t~Hoen, Monlong, Rivas, Gonz{\`a}lez-Porta, Kurbatova, Griebel,
  Ferreira, et~al\mbox{.}}{Lappalainen
  et~al.}{2013}]{lappalainen2013transcriptome}
Lappalainen, T., Sammeth, M., Friedl{\"a}nder, M.~R., AC‘t~Hoen, P., Monlong,
  J., Rivas, M.~A., Gonz{\`a}lez-Porta, M., Kurbatova, N., Griebel, T.,
  Ferreira, P.~G., et~al. (2013).
\newblock Transcriptome and genome sequencing uncovers functional variation in
  humans.
\newblock {\em Nature} {\bf 501,} 506--511.

\bibitem[\protect\citeauthoryear{Lawrence, Huber, Pages, Aboyoun, Carlson,
  Gentleman, Morgan, and Carey}{Lawrence et~al.}{2013}]{lawrence2013software}
Lawrence, M., Huber, W., Pages, H., Aboyoun, P., Carlson, M., Gentleman, R.,
  Morgan, M.~T., and Carey, V.~J. (2013).
\newblock Software for computing and annotating genomic ranges.
\newblock {\em PLoS Comput Biol} {\bf 9,} e1003118.

\bibitem[\protect\citeauthoryear{Luedi, Dietrich, Weidman, Bosko, Jirtle, and
  Hartemink}{Luedi et~al.}{2007}]{luedi2007computational}
Luedi, P.~P., Dietrich, F.~S., Weidman, J.~R., Bosko, J.~M., Jirtle, R.~L., and
  Hartemink, A.~J. (2007).
\newblock Computational and experimental identification of novel human
  imprinted genes.
\newblock {\em Genome research} {\bf 17,} 1723--1730.

\bibitem[\protect\citeauthoryear{Morcos, Ge, Koka, Lam, Pokholok, Gunderson,
  Montpetit, Verlaan, and Pastinen}{Morcos et~al.}{2011}]{morcos2011genome}
Morcos, L., Ge, B., Koka, V., Lam, K.~C., Pokholok, D.~K., Gunderson, K.~L.,
  Montpetit, A., Verlaan, D.~J., and Pastinen, T. (2011).
\newblock Genome-wide assessment of imprinted expression in human cells.
\newblock {\em Genome biology} {\bf 12,} 1.

\bibitem[\protect\citeauthoryear{Morison, Ramsay, and Spencer}{Morison
  et~al.}{2005}]{morison2005census}
Morison, I.~M., Ramsay, J.~P., and Spencer, H.~G. (2005).
\newblock A census of mammalian imprinting.
\newblock {\em TRENDS in Genetics} {\bf 21,} 457--465.
\newblock \url{http://www.otago.ac.nz/IGC}.

\bibitem[\protect\citeauthoryear{Peters}{Peters}{2014}]{peters2014role}
Peters, J. (2014).
\newblock The role of genomic imprinting in biology and disease: an expanding
  view.
\newblock {\em Nature Reviews Genetics} {\bf 15,} 517--530.

\bibitem[\protect\citeauthoryear{Rhead, Karolchik, Kuhn, Hinrichs, Zweig,
  Fujita, Diekhans, Smith, Rosenbloom, Raney, et~al\mbox{.}}{Rhead
  et~al.}{2009}]{rhead2009ucsc}
Rhead, B., Karolchik, D., Kuhn, R.~M., Hinrichs, A.~S., Zweig, A.~S., Fujita,
  P.~A., Diekhans, M., Smith, K.~E., Rosenbloom, K.~R., Raney, B.~J., et~al.
  (2009).
\newblock The ucsc genome browser database: update 2010.
\newblock {\em Nucleic acids research} page gkp939.

\bibitem[\protect\citeauthoryear{Sun}{Sun}{2012}]{sun2012asSeq}
Sun, W. (2012).
\newblock A statistical framework for {eQTL} mapping using {RNA}-seq data.
\newblock {\em Biometrics} {\bf 68,} 1--11.

\bibitem[\protect\citeauthoryear{Sun and Hu}{Sun and Hu}{2013}]{sun2013eqtl}
Sun, W. and Hu, Y. (2013).
\newblock eqtl mapping using rna-seq data.
\newblock {\em Statistics in biosciences} {\bf 5,} 198--219.

\bibitem[\protect\citeauthoryear{Trapnell, Pachter, and Salzberg}{Trapnell
  et~al.}{2009}]{trapnell2009tophat}
Trapnell, C., Pachter, L., and Salzberg, S.~L. (2009).
\newblock Tophat: discovering splice junctions with {RNA}-seq.
\newblock {\em Bioinformatics} {\bf 25,} 1105--1111.

\bibitem[\protect\citeauthoryear{Zou, Sun, Crowley, Zhabotynsky, Sullivan, and
  de~Villena}{Zou et~al.}{2014}]{zou2014novel}
Zou, F., Sun, W., Crowley, J.~J., Zhabotynsky, V., Sullivan, P.~F., and
  de~Villena, F. P.-M. (2014).
\newblock A novel statistical approach for jointly analyzing {RNA}-seq data
  from f1 reciprocal crosses and inbred lines.
\newblock {\em Genetics} {\bf 197,} 389--399.

\end{thebibliography}

\label{lastpage}

\end{document}